\documentclass[aps,twocolumn,superscriptaddress,PRD]{revtex4}
\usepackage{amsmath}
\usepackage{epsfig}
\usepackage{multirow}
\usepackage{slashed}
\usepackage{amsmath}
\usepackage{color}
\usepackage{float}
\usepackage{graphicx}
\usepackage[colorlinks,
            linkcolor=blue,
            anchorcolor=blue,
            citecolor=blue
            ]{hyperref}

\def\bea#1\eea{\begin{align}#1\end{align}}

\newcommand{\bef}{\begin{figure}[!htp]}
\newcommand{\eef}{\end{figure}}

\begin{document}
\title{Multiprocess imaging of nuclear modifications on parton distributions
in proton-nucleus collisions}

\author{Meng-Quan Yang}
\affiliation{Key Laboratory of Quark $\&$ Lepton Physics (MOE) and Institute of Particle Physics, Central China Normal University, Wuhan 430079, China}

\author{Peng Ru}
\email{p.ru@m.scnu.edu.cn}
\affiliation{School of Materials and New Energy, South China Normal University, Shanwei 516699, China}

\affiliation{State Key Laboratory of Nuclear Physics and Technology, Institute of Quantum Matter, South China Normal University, Guangzhou 510006, China}

\affiliation{Guangdong Basic Research Center of Excellence for Structure and Fundamental Interactions of Matter, Guangdong Provincial Key Laboratory of Nuclear Science, Guangzhou 510006, China}

\author{Ben-Wei Zhang}
\email{bwzhang@mail.ccnu.edu.cn}
\affiliation{Key Laboratory of Quark $\&$ Lepton Physics (MOE) and Institute of Particle Physics, Central China Normal University, Wuhan 430079, China}

\date{\today}

\begin{abstract}
Nuclear modifications to collinear parton distribution functions are conventionally quantified by the ratios \( r^{\textrm{A}}_i(x,Q^2) = f^\textrm{A,proton}_i(x,Q^2) / f^\textrm{proton}_i(x,Q^2) \). For a given nucleus $A$, these ratios generally depend on the parton momentum fraction \( x \), the probing scale \( Q^{2} \), and the parton species \( i \). Determining these dependencies relies on a global analysis of diverse experimental data. However, in realistic observables, these dependencies are intricately intertwined, making their extraction challenging. In this paper, we propose a novel approach to effectively image the nuclear modification factors \( r^{\textrm{A}}_i(x,Q^2) \) at the observable level in proton-nucleus collisions at the Large Hadron Collider. Specifically, through a combined study of \( Z \)-boson production, \( Z \)+jet production, and \( Z+c \)-jet production, we separately enhance signals arising from light-quark, gluon, and heavy-flavor (charm) distributions in nuclei. This enables us to effectively image the \( r^{\textrm{A}}_i(x,Q^2) \) for specific parton species. The feasibility of this method is validated through perturbative calculations at next-to-leading order in the strong coupling constant, employing three sets of nuclear PDF parametrizations: EPPS21, nCTEQ15, and TUJU19. Future measurements of these observables are expected to provide better-motivated parametrization form of nuclear PDFs and yield new insights into the detailed partonic structures of nuclei.

\end{abstract}

\maketitle

\section{Introduction}
Understanding the partonic structures within a nucleon and their modifications in a nucleus is fundamental to the study of quantum chromodynamics~(QCD)~\cite{Kovarik:2019xvh,Ethier:2020way,Gross:2022hyw,Klasen:2023uqj} and represents a key objective of upcoming experiments at the Electron-Ion Collider~(EIC)~\cite{AbdulKhalek:2021gbh,Anderle:2021wcy,Armesto:2023hnw}. In the phenomenology of high-energy nuclear collisions, a critical aspect is the nuclear modifications to collinear parton distribution functions~(PDFs)~\cite{Klasen:2023uqj}, which provide an essential baseline for disentangling final-state nuclear matter effects probed by hard particles~\cite{Gyulassy:2003mc,Qin:2015srf,Cao:2020wlm,Cunqueiro:2021wls,Xie:2024xbn,Cao:2024pxc,Mehtar-Tani:2024jtd, Yang:2023dwc, Andres:2024hdd}. These modifications are typically quantified by the ratios of nuclear PDFs to free-nucleon PDFs, expressed as $r^{\textrm{A}}_i(x,Q^2)\!=\!f^\textrm{A,proton}_i(x,Q^2)/f^\textrm{proton}_i(x,Q^2)$. These ratios are primarily extracted from global analyses of both the denominators and numerators and generally depend on the parton momentum fraction $x$, the probing scale $Q^{2}$, and the parton species $i$~\cite{Klasen:2024xqn,Dulat:2015mca,Segarra:2020gtj,Hirai:2007sx,Eskola:2009uj,deFlorian:2011fp,Kovarik:2015cma,Wang:2016mzo,Walt:2019slu,Khanpour:2020zyu,Eskola:2021nhw,AbdulKhalek:2022fyi}.

The connection between PDFs, their nuclear modifications, and observables is established through collinear factorization in perturbative QCD~\cite{Kovarik:2019xvh,EllisQCD}. However, the dependencies on $x$, $Q^2$, and $i$ are intricately convoluted in calculations, making it challenging to infer the explicit form of $r^{\textrm{A}}_i(x,Q^2)$ directly from experimental data~\cite{Klasen:2023uqj,Klasen:2024xqn}. Insights from nuclear binding dynamics, such as off-shell corrections and short-range correlations, could significantly improve the extraction of nuclear PDFs and the $r^{\textrm{A}}_i(x,Q^2)$ factors~\cite{Alekhin:2022tip,Alekhin:2022uwc,Cloet:2012td,CLAS:2019vsb,Cocuzza:2021rfn,Dalal:2022zkg,nCTEQ:2023cpo}. Additionally, specific observables, such as nuclear modifications to structure functions in deep-inelastic scattering~(DIS), can serve as effective images of $r^{\textrm{A}}_i(x,Q^2)$ for quark distributions~\cite{EuropeanMuon:1981gyc,Bodek:1983qn,Arneodo:1992wf,Seely:2009gt}. The measurements of such observables often inspire parametrizations of modifications, typically categorized into shadowing, anti-shadowing, EMC, and Fermi-motion regions across different $x$ ranges~\cite{Armesto:2006ph}, which greatly assist global analyses.

\begin{figure*}[t]
\hspace{-0.5cm}\includegraphics[width=2.3in]{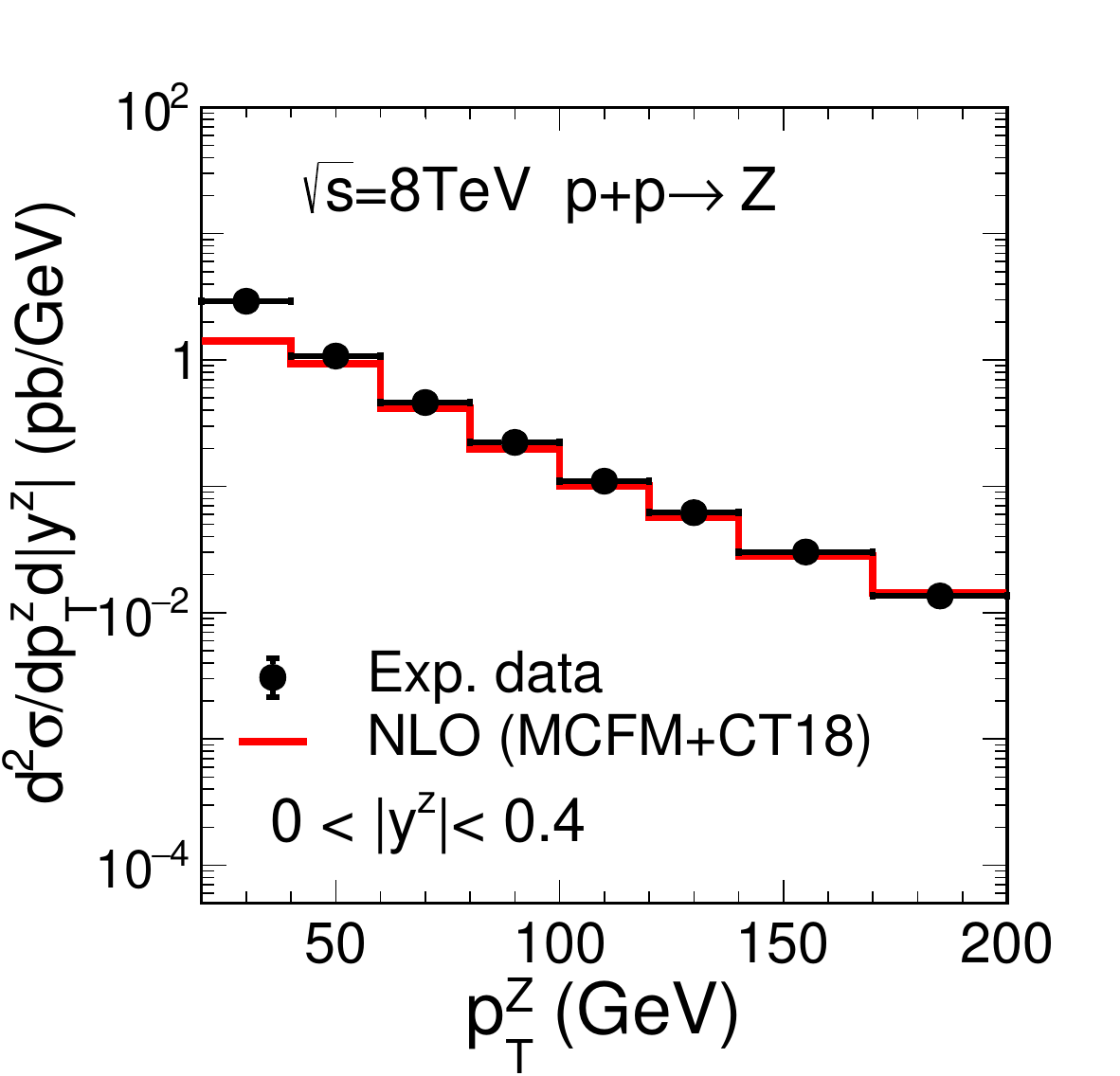}
\includegraphics[width=2.3in]{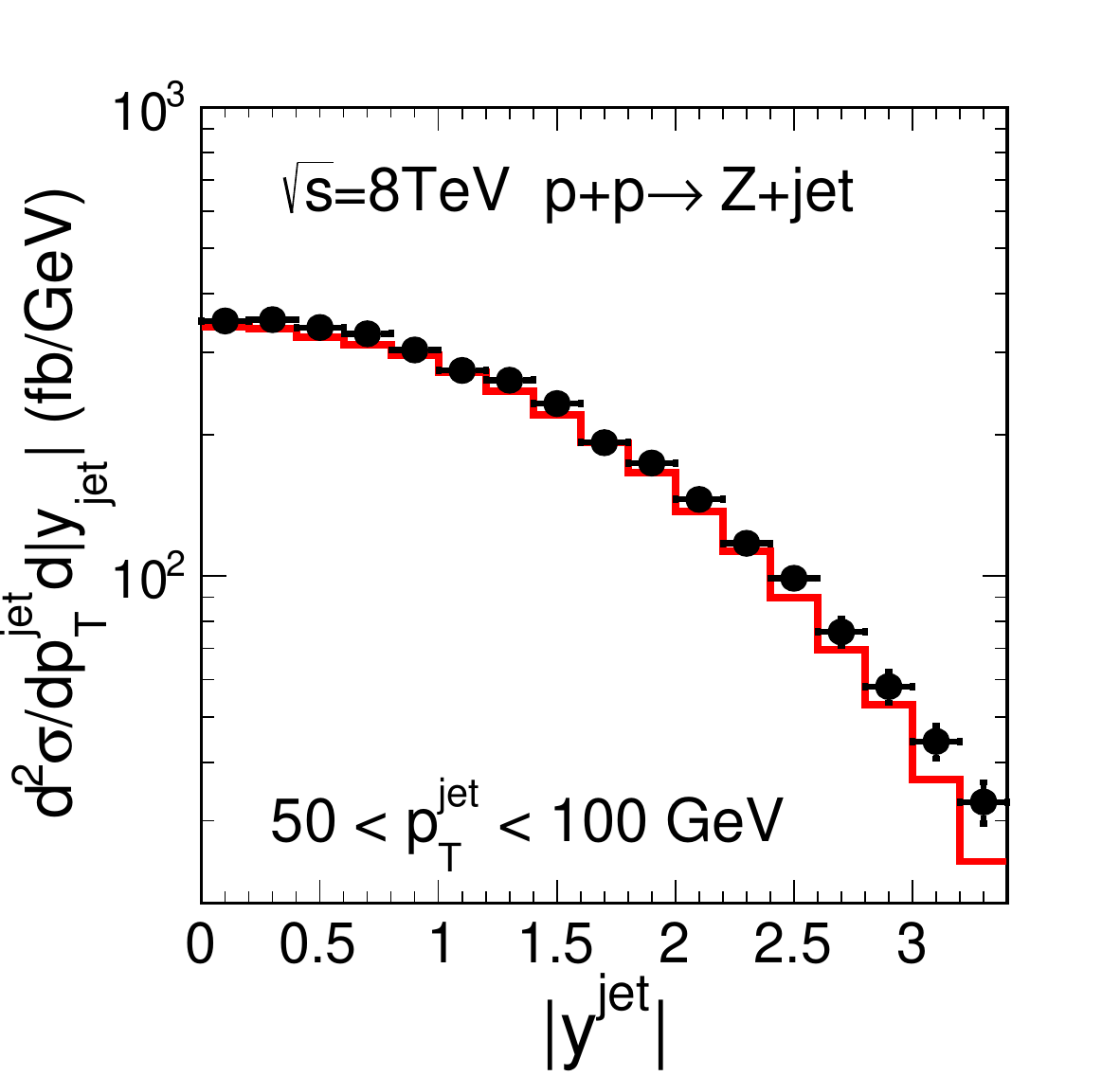}
\includegraphics[width=2.3in]{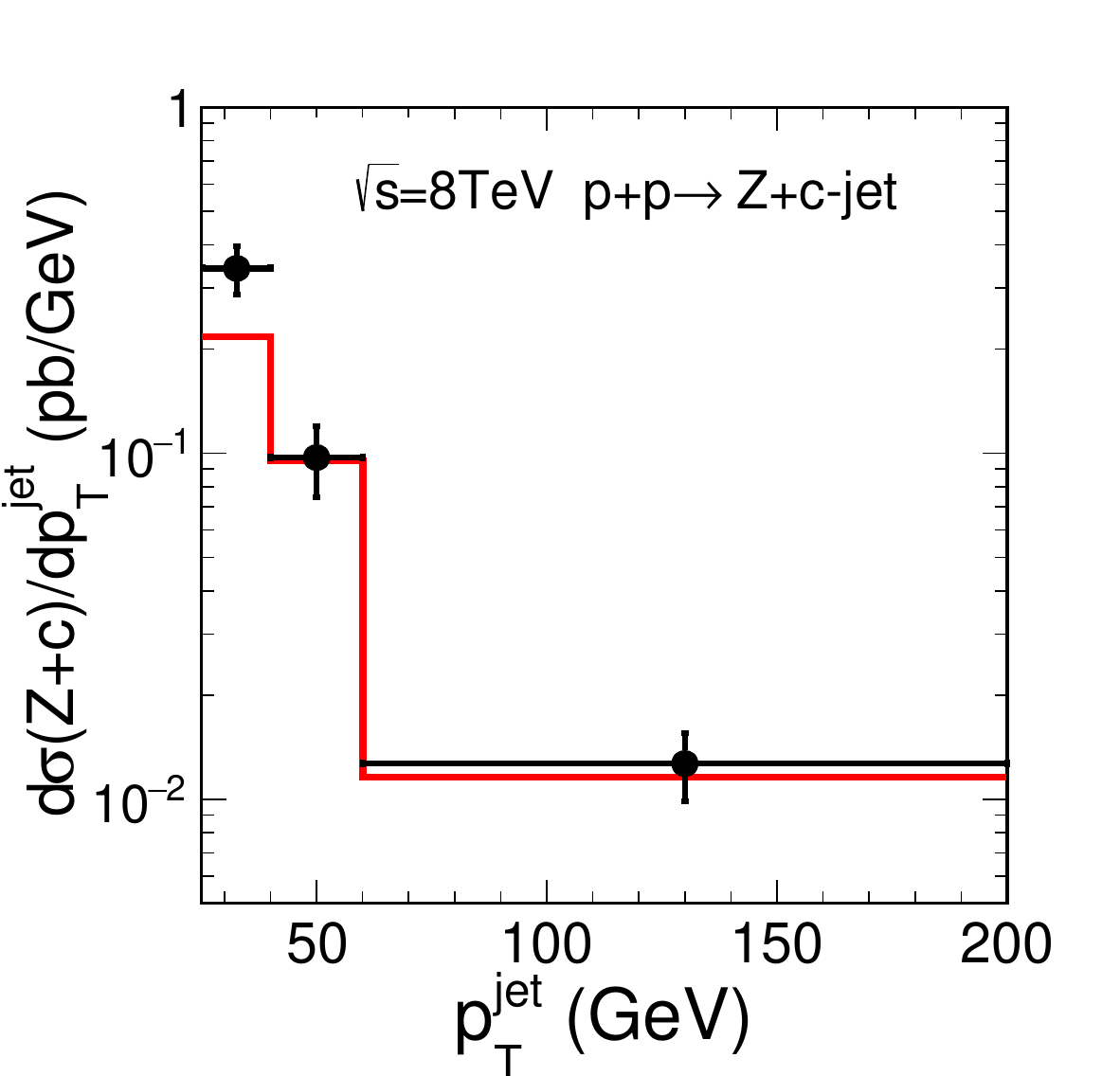}
\caption{Typical measurements of differential cross sections for $Z$-boson production~(left panel), $Z\!+\!$jet production~(middle panel), and $Z\!+\!c\!-\!$jet production~(right panel) in $pp$ collisions at $\sqrt{s}=8$~TeV at the LHC. Measurements are binned according to the rapidity $y^Z$ and transverse momentum $p_T^Z$ for $Z$-boson production~\cite{CMS:2015hyl}, the rapidity $y^{\textrm{jet}}$ and transverse momentum $p_T^{\textrm{jet}}$ for $Z\!+\!$jet production~\cite{ATLAS:2019bsa}, and the transverse momentum $p_T^{\textrm{jet}}$ for $Z\!+\!c\!-\!$jet production~\cite{CMS:2017snu}. Experimental data are represented by black discs, while NLO predictions computed with MCFM~\cite{Campbell:1999ah,Campbell:2011bn,Campbell:2015qma,Campbell:2021vlt} are depicted by red lines.}
\label{fig:pp_mcfm}
\end{figure*}

The vast amount of data from the Large Hadron Collider~(LHC) has revolutionized global analyses over the past decade~\cite{Klasen:2023uqj,Klasen:2024xqn}, leading to more stringent constraints on nuclear PDFs, notably for gluon distributions~\cite{Kusina:2017gkz,Eskola:2019dui,Eskola:2019bgf,Kusina:2020dki,Paakkinen:2022qxn}, and advancing neural-network-based analyses that reduce reliance on specific parametrizations~\cite{AbdulKhalek:2022fyi}. However, traditional measurements in proton-proton~($pp$) and proton-nucleus~($p$A) collisions exhibit a less direct mapping to PDFs, and the imaging of $r^{\textrm{A}}_i(x,Q^2)$ achieved in DIS has not been replicated for most LHC processes~\cite{ALICE:2013snh,LHCb:2017ygo,CMS:2018jpl,ALICE:2023ama,ATLAS:2024mvt}. This makes systematic comparisons of nuclear modifications across different processes challenging. Furthermore, significant discrepancies among widely used nuclear PDF parametrizations in certain regions remain unresolved~\cite{Klasen:2023uqj,Klasen:2024xqn}.

In a previous study~\cite{Shen:2021eir}, a method to reorganize the differential cross section for dijet production, initially proposed by Ellis and Soper~\cite{Ellis:1994dg}, was extended to $p$A collisions. This approach enables a well-controlled kinematic scan over $x$ and $Q^2$ for nuclear PDFs, providing an overall image of the $r^{\textrm{A}}_i(x,Q^2)$ factors~\cite{Shen:2021eir}. However, this image combines contributions from different parton species simultaneously. Given that the relative contributions of various parton species vary across processes, extending this method to other $p$A collision processes is of significant interest. Specifically, once kinematic scans are established for these processes, it becomes possible to enhance signals for specific parton flavors by combining results from different processes.

In this paper, we extend the method from Ref.~\cite{Shen:2021eir} to $Z$-boson, $Z+$jet, and $Z+$charm-jet production in proton-lead~($p$Pb) collisions at the LHC, enabling kinematic scans of nuclear PDFs for each process. The partonic subprocesses at leading order (LO) are listed in Tab.~\ref{tab:subprocess}. 
\begin{table}[h]
    \centering
    \begin{tabular}{c|c}
       \hline\
       \,\,\,Process\,\,\,  & \,\,\,partonic subprocess at LO\,\,\,\\\hline
       $Z$-boson  & $q+\bar{q}\rightarrow Z\rightarrow l^+l^-$ \\ \hline 
        $Z+$jet & $q+\bar{q}\rightarrow Z+g$ \\
         & $q(\bar{q})+g\rightarrow Z+q(\bar{q})$ \\\hline
        $Z+c$-jet & $c(\bar{c})+g\rightarrow Z+c(\bar{c})$\\\hline
    \end{tabular}
    \caption{Partonic subprocesses at LO for $Z$-boson production, $Z+$jet production, and $Z+$charm-jet production, respectively.}
    \label{tab:subprocess}
\end{table}
By further combining the differential cross sections of these processes, we design observables specifically sensitive to light quark, gluon, and heavy-flavor (charm) distributions in nuclei. Using next-to-leading order~(NLO) perturbative QCD calculations with three nuclear PDF parametrizations—EPPS21 \cite{Eskola:2021nhw}, nCTEQ15 \cite{Kovarik:2015cma}, and TUJU19 \cite{Walt:2019slu}—we demonstrate that the nuclear modifications to the proposed (combined) cross sections can effectively image the $r^{\textrm{A}}_i(x,Q^2)$ for specific parton species.

The paper is organized as follows: In Sec.~\ref{sec:baseline}, we extend the method from Ref.~\cite{Shen:2021eir} to reorganize the differential cross sections for $Z$-boson, $Z+$jet, and $Z+$charm-jet production in $p$Pb collisions. We then present an imaging of $r^{\textrm{A}}_i(x,Q^2)$ for light quarks using $Z$-boson production. In Sec.~\ref{sec:multi-process}, we design two combinations of multi-process cross sections and show that their nuclear modifications can image $r^{\textrm{A}}_i(x,Q^2)$ for gluon and heavy-flavor (charm) distributions, respectively. A summary and discussion are provided in Sec.~\ref{sec:summary}.

\begin{figure*}[t]
\hspace{-0.5cm}\includegraphics[width=2.5in]{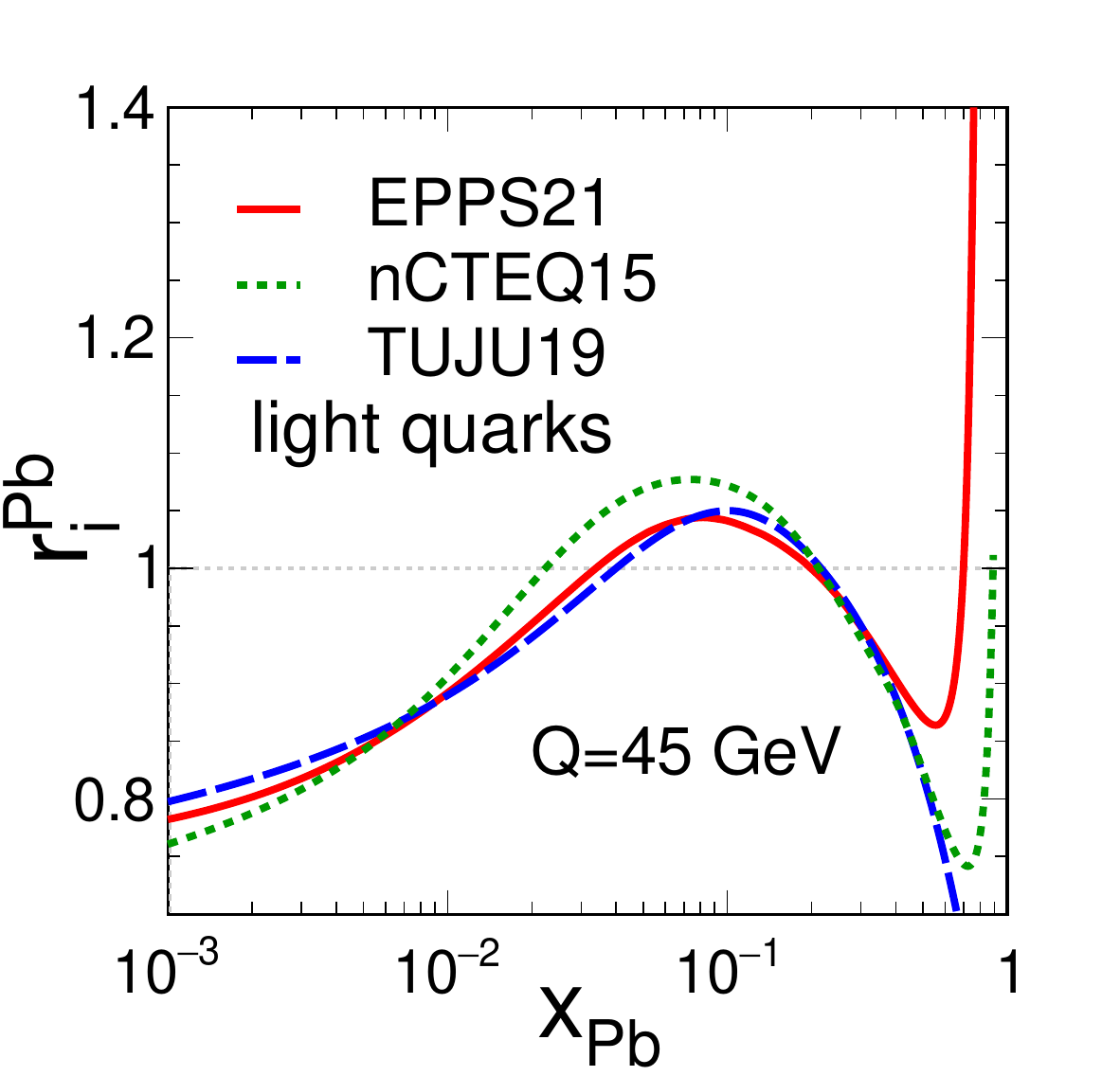}
\hspace{-0.5cm}\includegraphics[width=2.5in]{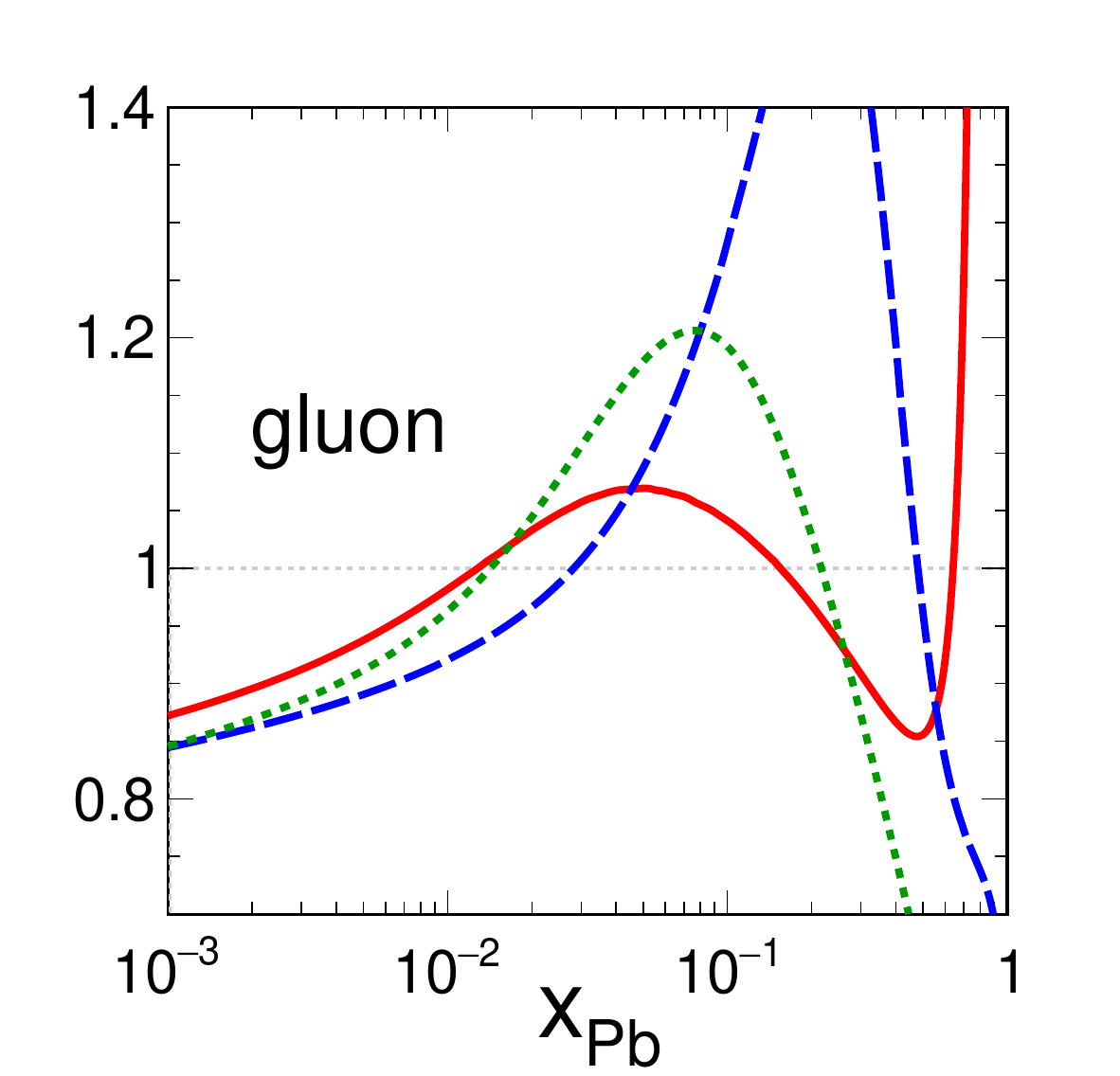}
\hspace{-0.5cm}\includegraphics[width=2.5in]{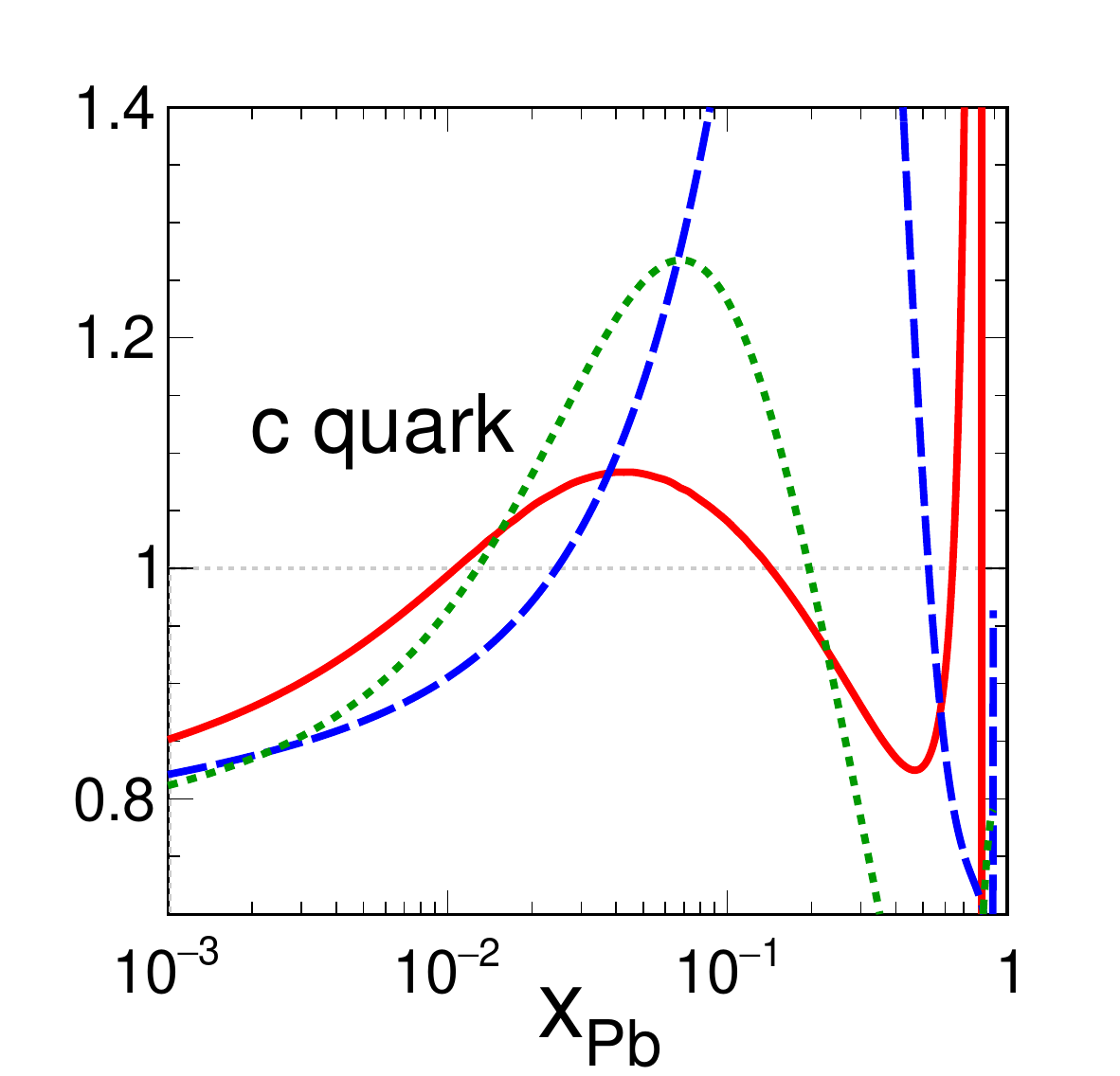}
\caption{Nuclear correction factors $r_i^{Pb}(x, Q^2)$ for light-quark~(left panel), gluon~(middle panel), and charm quark~(right panel) distributions, as provided by the EPPS21, nCTEQ15, and TUJU19 nuclear PDFs at $Q\!=\!M_Z/2\!=\!45$~GeV. Light quark distribution is defined as $f_q(x)=\sum_{i=u,d,s}[f_i(x)+\bar{f}_i(x)]$.}
\label{fig:r_npdf}
\end{figure*}
\section{Reorganizing differential cross section in proton-nucleus collisions}
\label{sec:baseline}
Traditional measurements in $pp$ and $p$A collisions are typically conducted in the form of differential cross sections, often double-differential, binned with variables such as the transverse momentum, rapidity, and invariant mass of the final-state particles~\cite{Albacete:2013ei,Albacete:2017qng}. For instance, illustrated in Fig.~\ref{fig:pp_mcfm} are three representative measurements for $Z$-boson production~\cite{CMS:2015hyl}, $Z+$jet production~\cite{ATLAS:2019bsa}, and $Z+$charm-jet production\cite{CMS:2017snu} in $pp$ collisions at the LHC, respectively. In general, such observables may not be ideal for probing the PDFs defined in terms of the parton momentum fraction $x$ and the probing scale $Q^2$. For example, in the case of $Z+$jet production, measuring the cross section binned with the jet transverse momentum $p_T^{\textrm{jet}}$ and the jet rapidity $y^{\textrm{jet}}$ (as shown in the middle panel of Fig.~\ref{fig:pp_mcfm}) only constrains the kinematics of the jet, while leaving the kinematics of the $Z$ boson unconstrained. Consequently, even at LO, neither the momentum fraction carried by the initial-state parton nor the physical scale of the process can be precisely deduced from these two measured variables alone.

A well-controlled kinematic scan of the PDFs can be achieved by reorganizing the measured variables to establish a correspondence with $x$ and $Q^2$ at LO~\cite{Shen:2021eir}. To constrain the momentum fraction carried by the initial-state nuclear parton in $p$A collisions (with the nucleus moving backward), a variable $X_B$ can be introduced~\cite{Shen:2021eir,Ellis:1994dg}, generally expressed as
\bea
X_{B}=\sum_{i} \frac{p_i^{-}}{\sqrt{s}}=\sum_{i} \frac{E_{Ti}}{\sqrt{s}}e^{- y_i}.
\label{eq:XB}
\eea
Here, the summation of the backward components of the light-cone momenta, $p_i^-=E_{Ti}e^{- y_i}$, is performed over all inclusively produced particles. Specifically, for $Z$-boson production with leptonic decay ($Z\rightarrow l^+l^-$), the summation includes the di-lepton, while for $Z+$jet or $Z\!+\!c$-jet production, it encompasses both the particles in the jet and the di-lepton from $Z$ decay. With this definition, $X_B$ precisely equals the momentum fraction $x_{\textrm{Pb}}$ of the parton from the lead nucleus in $p$Pb collisions at LO accuracy. However, this equivalence may not hold at NLO and beyond.

To further access the probing scale $Q^2$ for the scanning of PDFs, additional variables can be introduced in conjunction with $X_B$ to define the differential cross section~\cite{Shen:2021eir}. For $Z$-boson production, the invariant mass of the $Z$ boson (or the di-lepton), denoted as $M_V$, can be used to define a double differential cross section:
\bea
d\sigma/dX_BdM_V.
\label{eq:Zboson}
\eea
For $Z\!+\!(c)$jet production, which involves more complex dynamic channels, one can employ both the invariant mass of the $Z$-jet system, $M_{V\!J}$, and the averaged transverse momentum, $p_{T,avg}$, to define a triple differential cross section:
\bea
d\sigma/dX_BdM_{VJ}dp_{T,avg}.
\label{eq:Z-jet}
\eea

For such cross sections, the nuclear modification factor $R_{pA}$ as a function of $X_B$ is expressed as a ratio:
\bea
R^{Z+\textrm{jet}}_{pA}(X_B)=\frac{1}{A}\frac{d\sigma^{pA}/dX_B\,dM_{V\!J}\,dp_{T,avg}}{d\sigma^{pp}/dX_B\,dM_{V\!J}\,dp_{T,avg}},
\label{eq:rpa}
\eea
where the nuclear mass number $A$ serves as a normalization factor, and the variables $M_{V\!J}$ and $p_{T,avg}$ are fixed within specific bins to control the physical scales. This $R_{pA}$ factor is closely related to the nuclear modifications on PDFs, quantified by $r^{\textrm{A}}_i(x,Q^2)$, but remains largely insensitive to the underlying proton PDFs~\cite{Shen:2021eir}.

To evaluate the feasibility of the method proposed in this work, we perform perturbative calculations at NLO using the MCFM numerical program~\cite{Campbell:1999ah,Campbell:2011bn,Campbell:2015qma,Campbell:2021vlt}. The solid curves in Fig.~\ref{fig:pp_mcfm} represent the theoretical predictions for the three processes in $pp$ collisions, demonstrating an overall good agreement with the experimental data. While the predicted production yields could be further refined by calculations beyond NLO, the $R_{pA}$ factors, defined as ratios, are expected to be less sensitive to higher-order corrections in general~\cite{Eskola:2009uj,Ru:2014yma,Khanpour:2020zyu,Helenius:2021tof}.

In our calculations, the CT18 PDFs~\cite{Hou:2019efy} are used for the incoming proton, while the $r^{\textrm{A}}_i(x,Q^2)$ factors from three parametrizations—EPPS21 \cite{Eskola:2021nhw}, nCTEQ15 \cite{Kovarik:2015cma}, and TUJU19 \cite{Walt:2019slu}—are applied on top of the CT18 PDFs for the colliding nucleus. Figure \ref{fig:r_npdf} displays the $r^{\textrm{A}}_i(x,Q^2)$ factors at $Q\!=\!M_Z/2\!\approx\!45$~GeV from the three nuclear PDF sets for light quarks (left panel), gluons (middle panel), and charm quarks (right panel), respectively. Notable distinctions can be observed in the factors for different parton species and parametrizations. It should be noted that the latest versions of nCTEQ15~\cite{Muzakka:2022wey,nCTEQ:2023cpo} and TUJU19~\cite{Helenius:2021tof} have not been used in this study, which does not compromise or hinder our goal of testing the sensitivity of the proposed observables.

\begin{figure*}[t]
\hspace{-0.5cm}
\includegraphics[width=7.0in]{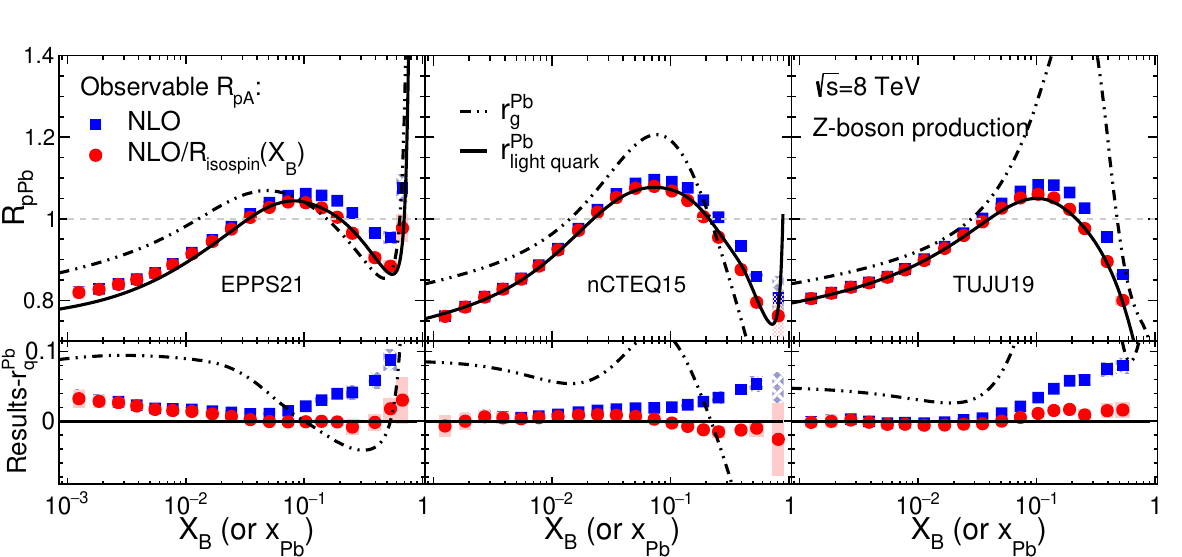}
\caption{Nuclear modification factors $R_{pA}$ as functions of $X_B$, calculated at NLO for reorganized cross section of $Z$-boson production (blue squares). Results corrected for isospin effects are represented by red discs. Predictions using EPPS21, nCTEQ15, and TUJU19 nuclear PDFs are displayed in left, middle, and right panels, respectively. Underlying $r^{\textrm{A}}_i(x,Q^2)$ factors for light quarks and gluons from corresponding nuclear PDFs are included in each panel for comparison. Bottom panels illustrate differences between each result and $r^{\textrm{A}}_i$ for light quarks. NLO calculations are conducted at scale $\mu_{0} = M_{Z}/2$. Uncertainties in the results, evaluated by varying scale between $\mu_{0}/2$ and $2\mu_{0}$, are represented by shaded bands.}
\label{fig:z_lq}
\end{figure*}

As an initial test, we investigate the $R_{pA}$ factor for the reorganized differential cross section of $Z$-boson production in $p$Pb collisions, as defined in Eq.~(\ref{eq:Zboson}). Since the partonic processes at LO are initiated by a $q\bar{q}$ pair, the $R_{pA}$ factors are expected to primarily reflect the nuclear modifications on the light-quark distributions. In the three panels of Fig.~\ref{fig:z_lq}, the NLO results for $R_{pA}$ are plotted as a function of $X_B$ (blue squares) using three sets of nuclear PDFs. The variable $M_V$ is constrained around the $Z$-mass peak, specifically within the range $66 < M_V < 116$~GeV. To contextualize these modifications, the $r^{\textrm{A}}_i(x,Q^2)$ curves for light-quark (solid) and gluon (dotted-dashed) distributions from the corresponding nuclear PDFs are also plotted in each panel of Fig.~\ref{fig:z_lq} as a function of $x_{\textrm{Pb}}$, with the probing scale $Q$ set at $M_Z/2$.

It is immediately apparent that the $R_{pA}$ values (blue squares) align closely with the $r^{\textrm{A}}_i$ for light quarks across most of the kinematic region. However, at large $X_B$ values ($>0.1$), the $R_{pA}$ shows an excess, primarily due to an isospin effect. Specifically, the valence quark ratio $d/u$ in the nucleus is higher than that in the proton, and the stronger weak coupling of the $d$ quark leads to enhanced $Z$-boson production in nuclear collisions~\cite{Halzen:2013bqa}. This isospin effect can be quantified by setting all $r^{\textrm{A}}_i$ factors to unity in the calculation of $R_{pA}$, expressed schematically as
\bea
R_{\textrm{isospin}}(X_B)=\frac{1}{A}\frac{Z\times d\sigma^{pp}+N\times d\sigma^{pn}}{d\sigma^{pp}},
\label{eq:riso}
\eea
where $\sigma^{pp(n)}$ represents the cross section for proton-proton (neutron) collisions, and $Z(N)$ denotes the number of protons (neutrons) in the nucleus. By dividing the $R_{pA}$ by $R_{\textrm{isospin}}$, we obtain the isospin-corrected nuclear modifications, shown as red discs in each panel of Fig.~\ref{fig:z_lq}. These corrected results exhibit excellent agreement with the $r^{\textrm{A}}_i$ for light-quark distributions over a wide range of $x$. For a more detailed comparison, the bottom panels of Fig.~\ref{fig:z_lq} display the differences between each result and the corresponding $r^{\textrm{A}}_i$ for light quarks. It is evident that, for all three nuclear PDFs, the red discs lie on or near the horizon, indicating strong consistency.

For this analysis, we have set the factorization and renormalization scales~($\mu_F$ and $\mu_R$) in the NLO calculations as $\mu_F=\mu_R=\mu_0$, with $\mu_0=M_Z/2$. The uncertainty in $R_{pA}$, evaluated by varying both between $\mu_0/2$ and $2\mu_0$, is depicted as shaded bands in Fig.~\ref{fig:z_lq}. As anticipated, these variations are minimal across most of the kinematic region. In general, the $R_{pA}$ factors obtained at NLO reflect the underlying nuclear modifications to the PDFs at the hard-scattering-related resolution scale(s)~\cite{Shen:2021eir,Collins:1989gx}, which is linked to the variable(s) used to define the differential cross section, such as $M_V$ in this case.

\begin{figure*}[t]
\hspace{-0.5cm}\includegraphics[width=2.3in]{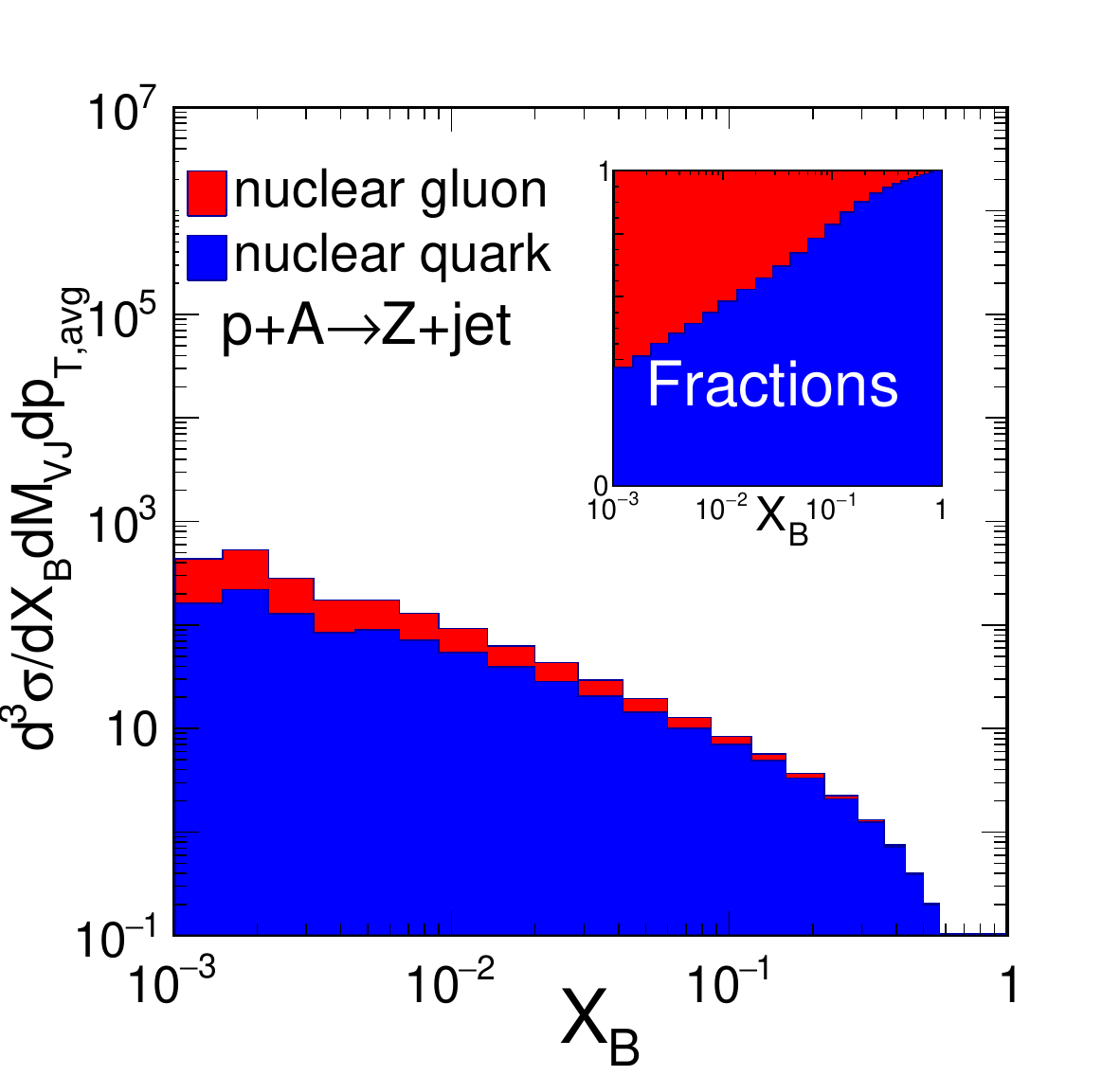}
\hspace{-0.5cm}\includegraphics[width=2.3in]{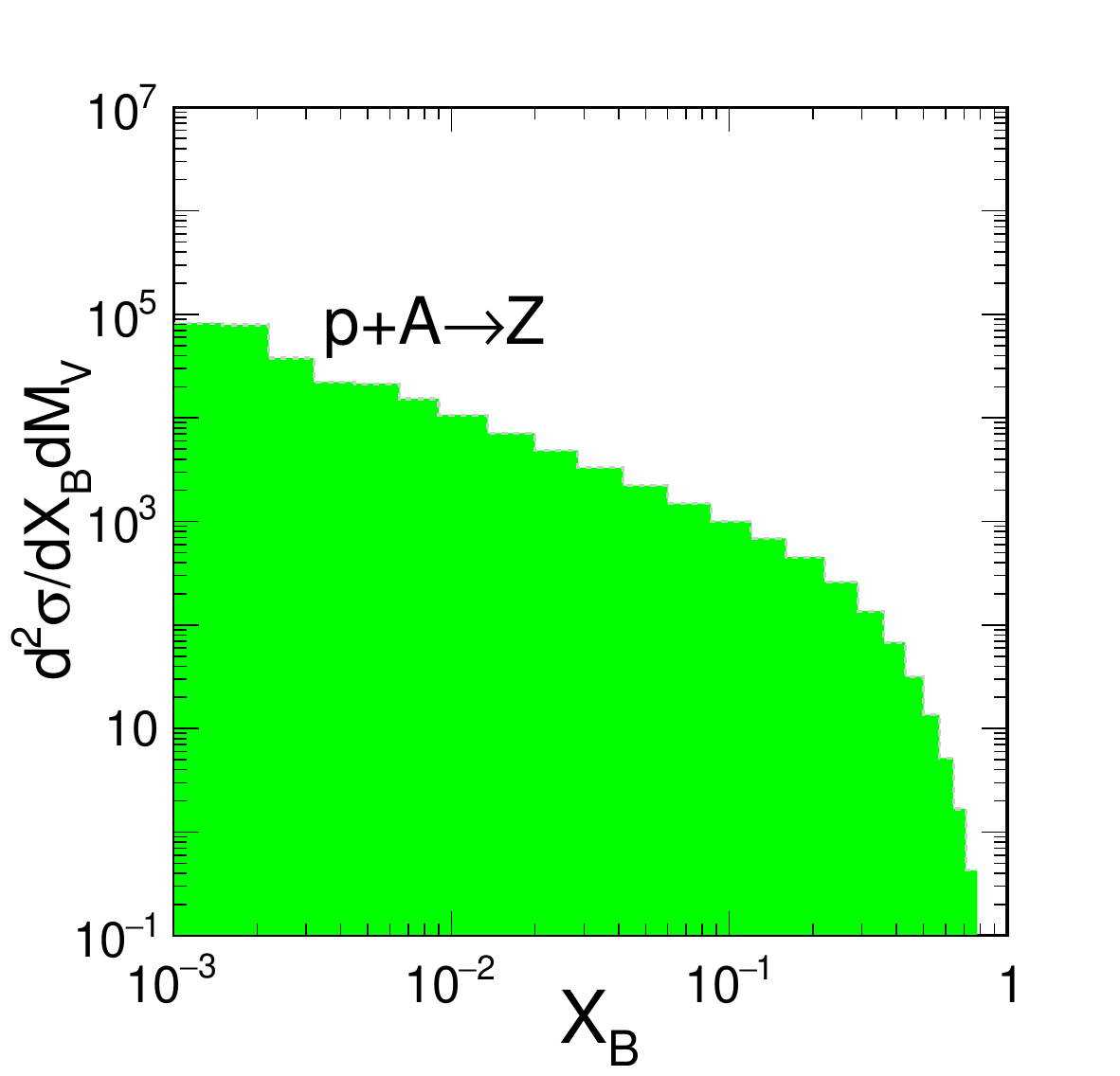}
\hspace{-0.5cm}\includegraphics[width=2.3in]{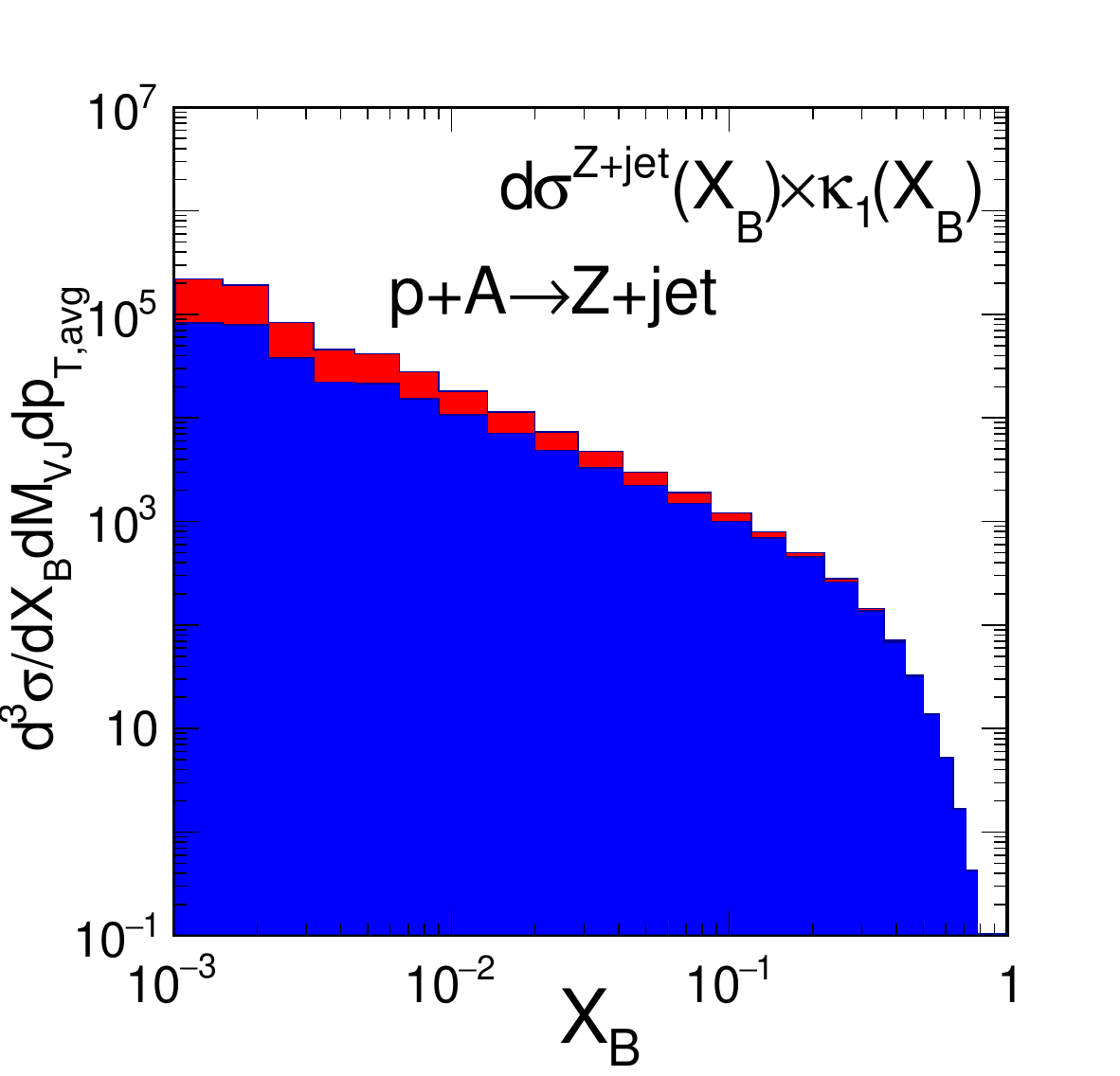}
\caption{Reorganized differential cross sections as functions of $X_B$, calculated at LO. Left panel: Results for $Z+$jet production, with contributions from nuclear quarks and gluons depicted as blue and red areas, respectively (inserted panel displays corresponding contribution fractions). Middle panel: Results for $Z$-boson production. Right panel: Rescaled results for $Z+$jet production, where nuclear quark contribution aligns with $Z$-boson cross section shown in middle panel. Calculations are performed at $\mu_F\!=\!\mu_R\!=\!\mu_0\!=\!M_Z/2$, using $pp$ collisions with backward-going proton treated as a nucleus.}
\label{fig:combine}
\end{figure*}

The study in this section indicates that the $R_{pA}$ factor for the reorganized differential cross section of $Z$-boson production offers an effective kinematic scan of the underlying nuclear modifications on the light quark distributions. However, it is important to note that this method may not be directly applicable to other processes. For instance, in the case of $W$-boson production~\cite{Ru:2016wfx,Ru:2014yma,Ru:2015pfa}, which is also sensitive to quark distributions, the final-state decay products include a neutrino, leading to potentially incomplete four-momentum information. A similar challenge arises in $t\bar{t}$ production~\cite{dEnterria:2015mgr,ATLAS:2024qdu}, which is sensitive to gluon distributions. In such processes, the variable $X_B$ cannot be reliably reconstructed in experimental measurements.

\section{A multi-process imaging of $r^{\textrm{A}}_i$}
\label{sec:multi-process}
In this section, we extend the method of reorganizing the cross section to the study of $Z+$jet and $Z\!+\!c\!-\!$jet production. By leveraging the kinematic scan of nuclear PDFs in these processes, it becomes possible to separately enhance the signals of $r^{\textrm{A}}_i$ for gluon and charm quark distributions through a multi-process joint analysis.

For $Z+$jet production, the reorganized differential cross section can be defined in terms of $X_B$, $M_{V\!J}$, and $p_{T,avg}$ as outlined in Eq.~(\ref{eq:Z-jet}). At LO, the initial state involves both quarks and gluons, meaning the $R_{pA}$ for such cross sections incorporates additional contributions from nuclear gluons compared to $Z$-boson production. In the left panel of Fig.~\ref{fig:combine}, we present an example of the LO results for the reorganized cross section of $Z+$jet production in $p$A collisions. The contributions from nuclear quarks and gluons are depicted as blue and red areas, respectively. The inserted percentage diagram clearly illustrates that the contributions from nuclear gluons increase as $X_B$ decreases, consistent with expectations.

To isolate the nuclear gluon signal from the cross section, we assume that, under specific kinematic constraints, the nuclear modifications in quark-initiated processes approximate the $R_{pA}$ for $Z$-boson production at the LO level. This assumption will be validated later. Given that the differential cross section for $Z$-boson production~(shown in the middle panel of Fig.~\ref{fig:combine}) differs from the quark-initiated contribution to the $Z+$jet cross section, we introduce a ratio between the two as follows
\bea
\kappa_{1}(X_B)\!=\!\frac{d\sigma^{Z}(X_B)}{\,\,\,\left[d\sigma^{Z\!+\!\textrm{jet}}(X_B)\right]_{\textrm{nuclear quark}}}
\label{eq:cs1}
\eea
By multiplying the whole $Z+$jet cross section by the coefficient $\kappa_{1}(X_B)$, we obtain a rescaled $Z+$jet cross section, as illustrated in the right panel of Fig.~\ref{fig:combine}. At LO, the nuclear-quark component of the rescaled cross section aligns with the $Z$-boson production cross section. Further, by defining a difference term as
\bea
C_{g}(X_B)\!=\!\kappa_{1}(X_B)\times d\sigma^{Z\!+\!\textrm{jet}}(X_B)-d\sigma^{Z}(X_B),
\label{eq:cg}
\eea
we anticipate that the influence of the nuclear quark distribution will be significantly mitigated, allowing the $R_{pA}$ derived from $C_{g}(X_B)$ to highlight an enhanced signal originating from the nuclear gluon distribution. 

The coefficient $\kappa_{1}(X_B)$, defined at LO as expressed in Eq.~(\ref{eq:cs1}), can be further refined by multiplying a correction factor expressed as
\bea
R_{\textrm{NLO}}(X_B)\!=\!\frac{d\sigma^{Z}_{\textrm{NLO}}(X_B)/d\sigma^{Z}_{\textrm{LO}}(X_B)}
{\,\,\,d\sigma^{Z\!+\!\textrm{jet}}_{\textrm{NLO}}(X_B)/d\sigma^{Z\!+\!\textrm{jet}}_{\textrm{LO}}(X_B)\,\,\,},
\label{eq:RNLO}
\eea
which takes into account the differences between the NLO K-factor effects in the two combined processes to some extent.

\begin{figure*}[t]
\hspace{-0.5cm}\includegraphics[width=7in]{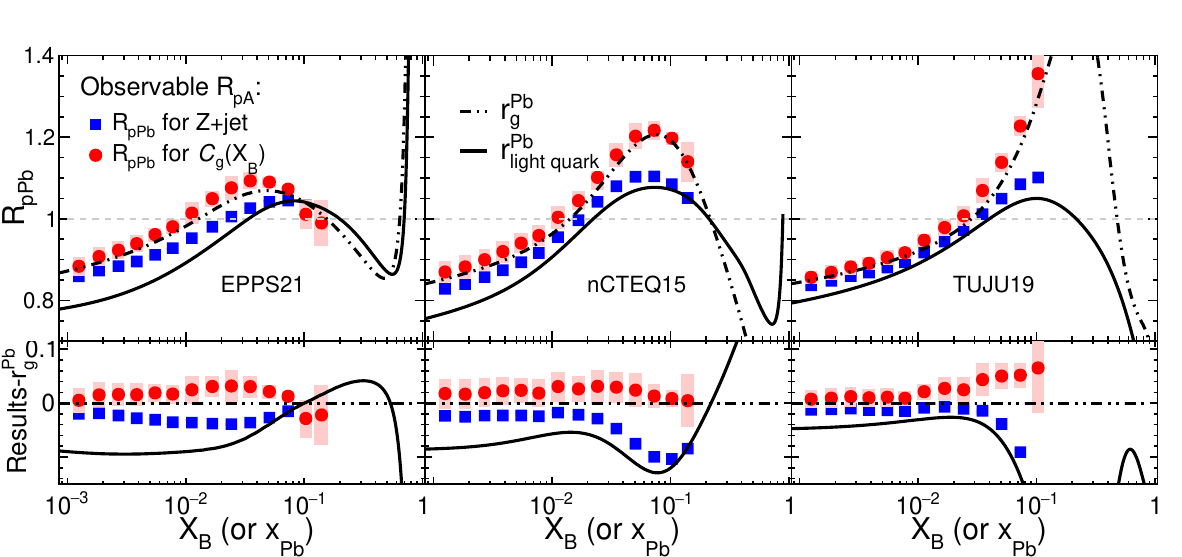}
\caption{Nuclear modification factors $R_{pA}$ as functions of $X_B$, calculated at NLO for reorganized cross section of $Z+$jet production (blue squares) and for combined observable $C_g(X_B)$~(red discs). Predictions using EPPS21, nCTEQ15, and TUJU19 nuclear PDFs are displayed in left, middle, and right panels, respectively. Underlying $r^{\textrm{A}}_i(x,Q^2)$ factors for light quarks and gluons from corresponding nuclear PDFs at $Q\!=\!M_Z/2$ are included in each panel for comparison. Bottom panels illustrate differences between each result and $r^{\textrm{A}}_i$ for gluon. NLO calculations are conducted at scale $\mu_{0} \!=\!M_{Z}/2$ for $Z$-boson and at $\mu_{0}\!=\!(M_{V\!J}\!+\!p_{T,avg})/4$ for $Z+$jet. Uncertainties in the results, evaluated by varying scale between $\mu_{0}/2$ and $2\mu_{0}$, are represented by shaded bands.}
\label{fig:zjet}
\end{figure*}

In the three panels of Fig.~\ref{fig:zjet}, the NLO results for the $R_{pA}$ of individual $Z+$jet production, as defined in Eq.~(\ref{eq:rpa}), are plotted as a function of $X_B$ (blue squares) using three sets of nuclear PDFs. By comparing these $R_{pA}$ values with the $r^{\textrm{A}}_i(x_{\textrm{Pb}})$ curves for light-quark (solid) and gluon (dotted-dashed) distributions in each panel, it is evident that the $R_{pA}$ squares generally lie between the quark and gluon curves. This reflects the fact that both nuclear parton species contribute significantly. Additionally, we present the corresponding NLO results for the $R_{pA}$ of $C_{g}(X_B)$, defined in Eq.~(\ref{eq:cg}), as red discs in each panel of Fig.~\ref{fig:zjet}. These results demonstrate that the $R_{pA}$ of $C_{g}$ closely follows the $r^{\textrm{A}}_i$ curves for the gluon distribution. For further comparison, the differences between each result and the corresponding $r^{\textrm{A}}_i$ for gluon are also plotted in the bottom panels of Fig.~\ref{fig:zjet}.

The above results are calculated for $Z+$jet production using the kinematic cuts $66 < M_{V\!J} < 150$~GeV and $40 < p_{T,avg} < 60$~GeV. These cuts ensure that the effective probing scales in this process are comparable to those in $Z$-boson production within the range $66 < M_{V} < 116$~GeV. As a test, we compute the average value of the quantity $(M_{V\!J} + p_{T,avg})/2$ for $Z+$jet production and find it to be close to the $Z$-boson mass $M_Z$. Additionally, in an LO test we observe that the $R_{pA}$ of $C_{g}$ aligns well with the $r^{\textrm{A}}_i$ for the gluon distribution, indicating that the nuclear-quark-initiated processes in $Z+$jet production experience modifications similar to those in $Z$-boson production.

\begin{figure*}[t]
\hspace{-0.5cm}\includegraphics[width=7in]{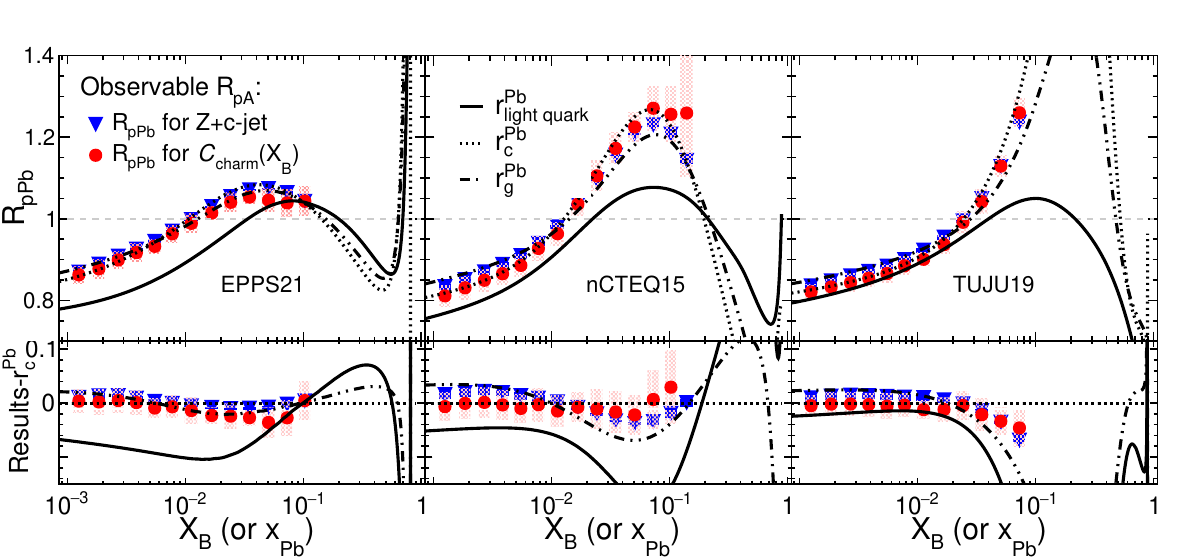}
\caption{Nuclear modification factors $R_{pA}$ as functions of $X_B$, calculated at NLO for reorganized cross section of $Z+\!c\!-\!$jet production (blue triangles) and for combined observable $C_{\textrm{charm}}(X_B)$~(red discs). Predictions using EPPS21, nCTEQ15, and TUJU19 nuclear PDFs are displayed in left, middle, and right panels, respectively. Underlying $r^{\textrm{A}}_i(x,Q^2)$ factors for light quarks, gluons and charm quarks from corresponding nuclear PDFs at $Q\!=\!M_Z/2$ are included in each panel for comparison. Bottom panels illustrate differences between each result and $r^{\textrm{A}}_i$ for charm quark. NLO calculations are conducted at scale $\mu_{0} \!=\!M_{Z}/2$ for $Z$-boson and at $\mu_{0}\!=\!(M_{V\!J}\!+\!p_{T,avg})/4$ for $Z+\!(c)$jet. Uncertainties in the results, evaluated by varying scale between $\mu_{0}/2$ and $2\mu_{0}$, are represented by shaded bands.}
\label{fig:zcjet}
\end{figure*}

In the calculations, the renormalization and factorization scales are set to $\mu_{0} = (M_{V\!J} + p_{T,avg})/4$ for $Z+$jet production, while for $Z$-boson production, they are set to $\mu_0 = M_Z/2$, consistent with the previous section. For $Z+$jet production, the variations in $R_{pA}$ with $\mu_{0}$ scaled by a factor of 2, represented as shaded bands around the squares in Fig.~\ref{fig:zjet}, are as small as those in $Z$-boson production. To illustrate the scale dependence for the $R_{pA}$ of the combined observable $C_{g}$, we assume that the choices of $\mu_{0}$ in the two processes are independent. Given that the ratio between $\mu^2_{0}$ in the two processes can vary by a factor of 16, the scale dependence of the combined observable generally appears stronger than that of a single observable, as shown in Fig.~\ref{fig:zjet}. The $R_{pA}$ of $C_{g}$ effectively images the $r^{\textrm{A}}_i$ for the gluon distribution up to a moderate value of $x_{\textrm{Pb}}$ ($\sim$0.1). For larger values of $x_{\textrm{Pb}}$, the uncertainties become significant, related to the rapidly diminishing contribution from gluons.

It should be mentioned that, as an input in the definition of $C_{g}(X_B)$, the coefficient $\kappa_{1}(X_B)$ itself depends on the theoretical setting in its determination. A possible impact of the free-proton PDF~(e.g.~scheme and parametrization) mainly arises from the relative compositions of various flavors in proton. To test the effects from proton PDF parametrization, we calculate the contribution fraction for certain nuclear parton flavors~(quark/gluon) with three parametrizations of free-proton PDFs~(CT18~\cite{Hou:2019efy}, MMHT~\cite{Harland-Lang:2019pla} and NNPDF~\cite{Cruz-Martinez:2024cbz}, see the Appendix~\ref{append:proton PDF}), and observe negligible distinctions among these results in most of the studied kinematic regions in this work. Besides, the PDF/factorization scheme may also have an impact. For example, at NLO and beyond the division between quark- and gluon-initiated channels is factorization-scheme dependent. Generally, the differences between the predictions with different schemes may become reduced when higher-order contributions are included. A quantitative study of the impact of the PDF scheme/factorization is beyond the scope of this work. Note that it is impossible to directly measure $r^{\textrm{A}}_i(x,Q^2)$ or completely purify the signal from a certain flavor. Therefore, the observable $C_{g}(X_B)$ is designed to establish an approximate and effective mapping to the underlying $r^{\textrm{A}}_i(x,Q^2)$ factors for gluon. In this work, the effectiveness of such observables is examined by NLO calculations with scale variations, and a nice reflection to the object $r^{\textrm{A}}_i(x,Q^2)$ is observed under the current scheme~~\cite{Campbell:1999ah,Campbell:2011bn,Campbell:2015qma,Campbell:2021vlt,Hou:2019efy}.

Following a similar methodology, we now investigate the production of $Z\!+\!c$-jet in $p$A collisions, a process that is sensitive to the nuclear modifications of both gluon and charm quark distributions.

\begin{figure*}[t]
\hspace{-0.5cm}\includegraphics[width=2.5in]{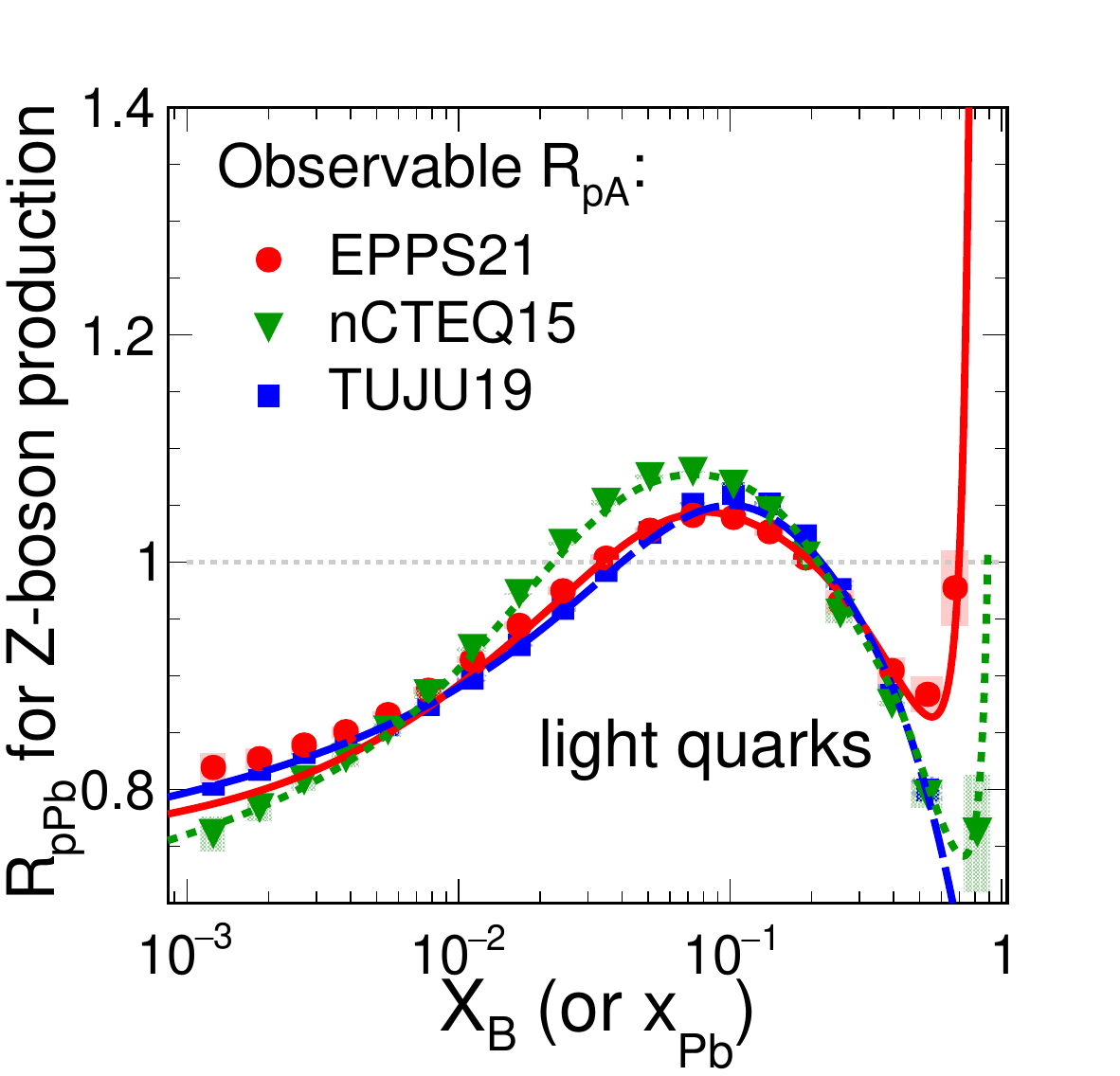}
\hspace{-0.5cm}\includegraphics[width=2.5in]{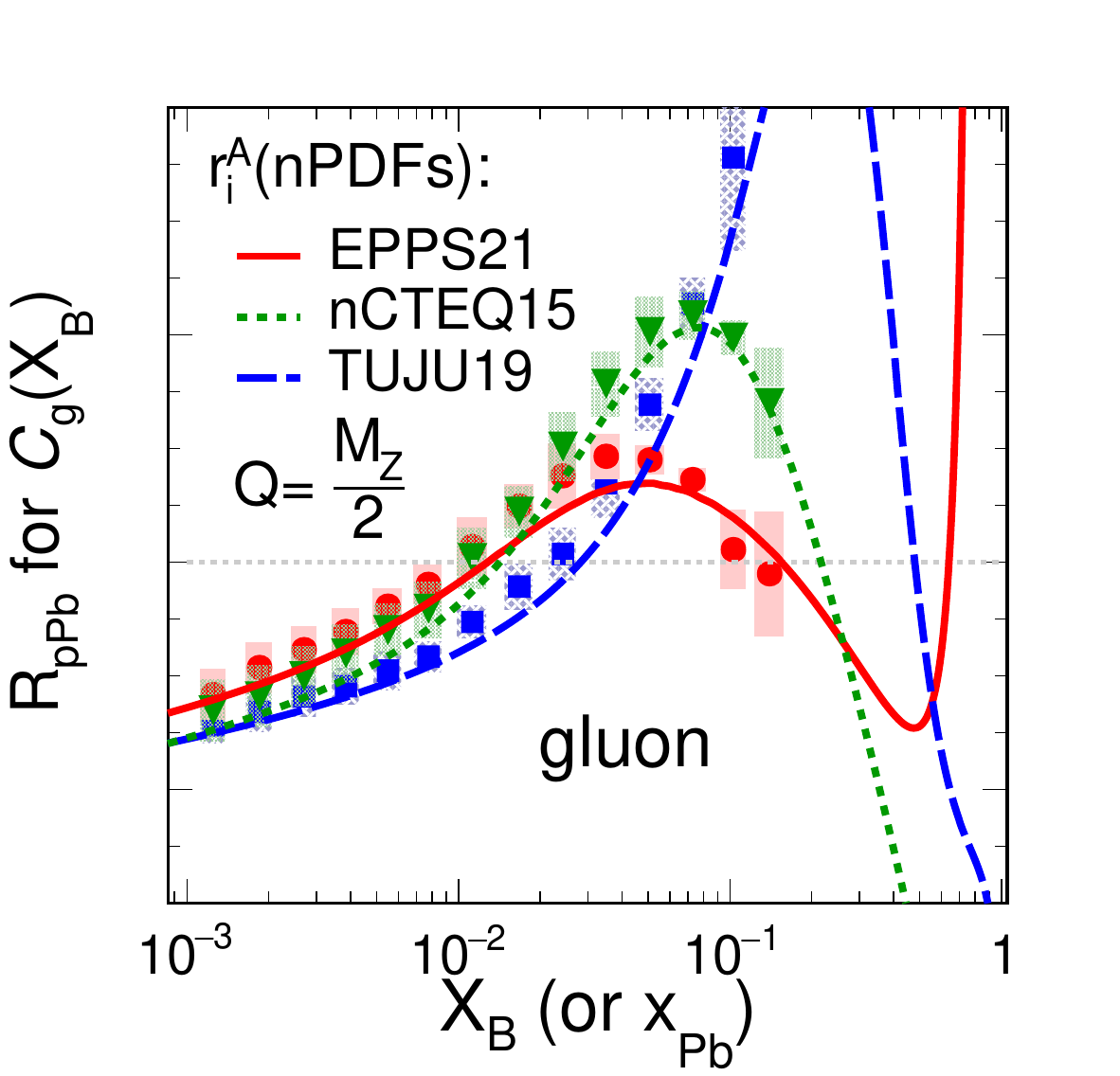}
\hspace{-0.5cm}\includegraphics[width=2.5in]{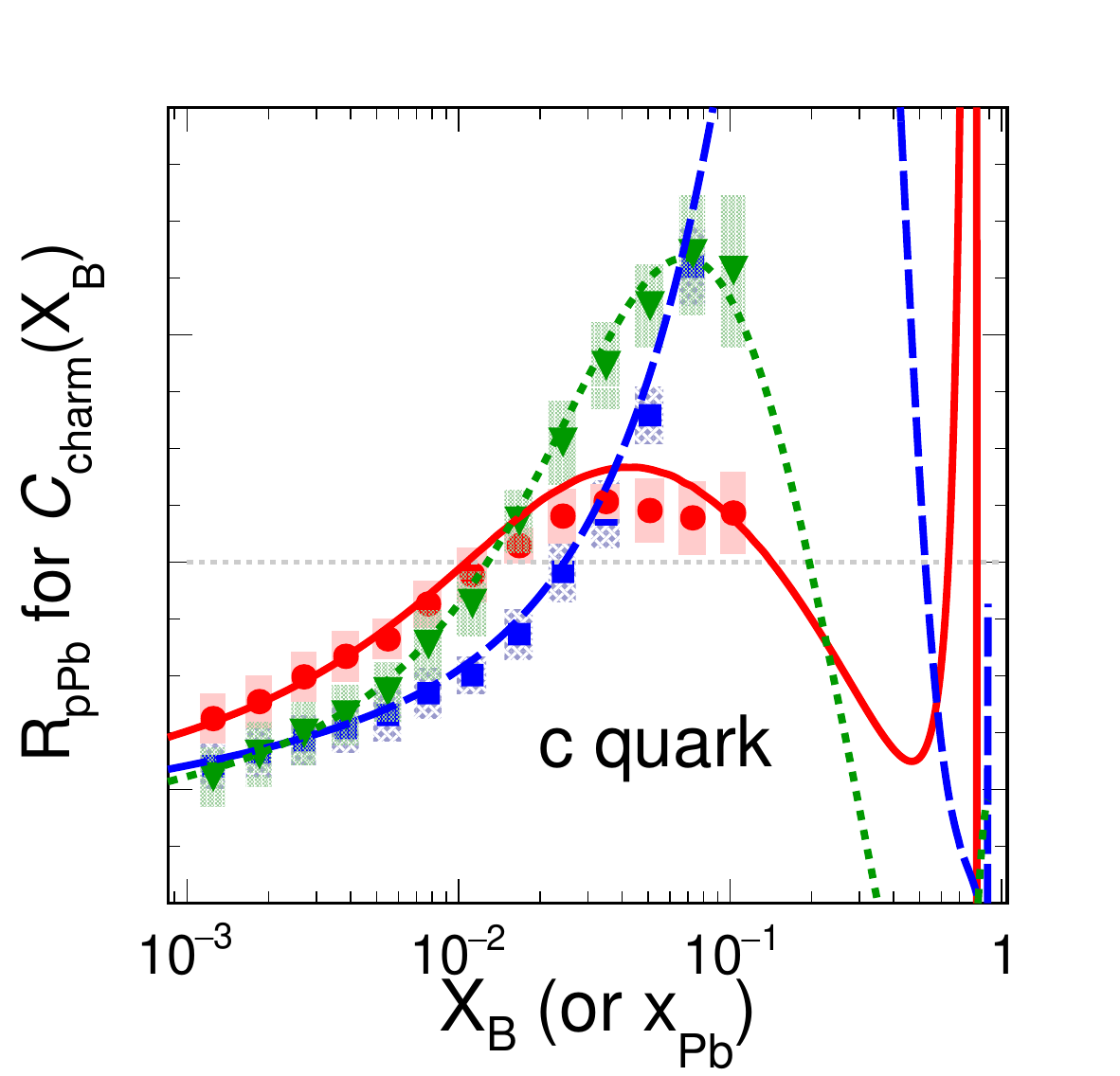}
\caption{Comparisons between predicted observable $R_{pA}$ and $r^{\textrm{A}}_i$ for three nuclear PDFs: EPPS21, nCTEQ15, and TUJU19. Left panel: $R_{pA}$ of reorganized cross section of $Z$-boson production~(isospin corrected) compared with $r^{\textrm{A}}_i$ for light quark. Middle panel: $R_{pA}$ of combined observable $C_g(X_B)$ compared with $r^{\textrm{A}}_i$ for gluon. Right panel: $R_{pA}$ of combined observable $C_{\textrm{charm}}(X_B)$ compared with $r^{\textrm{A}}_i$ for charm quark. 
}
\label{fig:image}
\end{figure*}

The $R_{pA}$ for the reorganized $Z\!+\!c$-jet cross section, as defined in Eq.~(\ref{eq:rpa}), is calculated at NLO using three sets of nuclear PDFs. These results are depicted as blue triangles in Fig.~\ref{fig:zcjet}, alongside the corresponding $r^{\textrm{A}}_i(x_{\textrm{Pb}})$ curves for light-quark (solid), gluon (dotted-dashed), and charm (dotted) distributions at $Q = M_Z/2$, included for comparative analysis. To maintain consistency in probing scales with the aforementioned processes, we employ kinematic cuts of $66 < M_{V\!J} < 150$~GeV and $20 < p_{T,avg} < 60$~GeV in the calculations of the reorganized $Z\!+\!c$-jet cross section. As illustrated in Fig.~\ref{fig:zcjet}, the $R_{pA}$ triangles closely align with the $r^{\textrm{A}}_i$ curves for both gluon and charm quark distributions, which are predicted to be similar across the three parametrizations. Additionally, the differences between each result and the corresponding $r^{\textrm{A}}_i$ for charm quark are presented in the bottom panels of Fig.~\ref{fig:zcjet} for further comparison.

Building on the study of the observable \( C_g(X_B) \), which is designed to probe the gluon distribution in nuclei, we can further develop a new combined multi-process observable specifically sensitive to the charm quark distribution. By iterating the procedures used to construct \( C_g(X_B) \), we define a coefficient
\bea
\kappa_{2}(X_B)\!=\!\frac{C_{g}(X_B)}{\,\,\,\left[d\sigma^{Z\!+\!c\!-\!\textrm{jet}}(X_B)\right]_{\textrm{nuclear gluon}}}
\label{eq:cs}
\eea
and a combined observable
\bea
C_{\textrm{charm}}(X_B)\!=\!\kappa_{2}(X_B)\times d\sigma^{Z\!+\!c\!-\!\textrm{jet}}(X_B)-C_{g}(X_B),
\label{eq:cc}
\eea
where the sensitivity to the gluon distribution is expected to be significantly reduced. It is important to note that the coefficients \(\kappa_{1}\) and \(\kappa_{2}\) serve as predetermined parameters in the definitions of the observables \(C_{g}\) and \(C_{\textrm{charm}}\). These coefficients are evaluated at a fixed scale \(\mu_0 = M_Z/2\) and remain invariant under changes in the renormalization and factorization scales in the NLO calculations for \(C_{g}\) or \(C_{\textrm{charm}}\).

Using the definition of \(C_{\textrm{charm}}(X_B)\), we calculate its nuclear modification factor \(R_{pA}(X_B)\) at NLO for three nuclear PDFs and present the results as red discs in Fig.~\ref{fig:zcjet}. As anticipated, since the gluon and charm distributions exhibit similar modifications in the three nuclear PDFs, the \(R_{pA}\) factors for the combined \(C_{\textrm{charm}}\) show only minor deviations from those for the individual \(Z\!+\!c\)-jet cross section (blue triangles). Moreover, they slightly better reproduce the \(r^{\textrm{A}}_i(x_{\textrm{Pb}})\) curves for the charm quark distribution. This demonstrates that the \(R_{pA}(X_B)\) of \(C_{\textrm{charm}}\) can effectively serve as an image of the \(r^{\textrm{A}}_i\) for the charm quark distribution.

To conclude this section, we consolidate the main findings of this work by presenting a comparison in Fig.~\ref{fig:image} between the proposed observables and the nuclear modifications on the underlying PDFs for light quark, gluon, and charm quark distributions. It is evident that the predicted $R_{pA}(X_B)$ at NLO using three nuclear PDFs aligns well with the corresponding $r^{\textrm{A}}_i(x_{\textrm{Pb}})$ for the three parton species: light quark, gluon, and heavy quark (charm). These results suggest that future measurements of such nuclear modifications have the potential to provide effective, observable-level images of the $r^{\textrm{A}}_i$ factors.

\section{SUMMARY AND DISCUSSION}
\label{sec:summary}
Nuclear modifications to parton distribution functions are primarily inferred through global analyses of a wide range of data within the framework of collinear factorization. However, determining the explicit dependence of the modification factors $r^{\textrm{A}}_i(x,Q^2)$ on the variables $x$, $Q^2$, and parton species $i$ remains challenging. This is partly due to their complex relationships with observables, the inherent uncertainties in both the data and predictions, and the parametrization dependence in global extractions~\cite{Klasen:2023uqj}. Besides continuous improvements in both experimental and theoretical precision, introducing observables with better disentangled contributions from PDF variables may help mitigate this situation.

In this work, we introduce a series of novel observables in proton-nucleus collisions at the LHC, designed to establish an approximate and effective mapping to the underlying nuclear modification factors $r^{\textrm{A}}_i(x,Q^2)$. Specifically, by combining the reorganized cross sections of $Z$-boson production, $Z$+jet production, and $Z+c$-jet production, we separately purify signals from light-quark, gluon, and heavy-flavor (charm) distributions in nuclei. This approach allows us to effectively image the $r^{\textrm{A}}_i(x,Q^2)$ for specific parton species, serving as an analogy to the measurement of nuclear modifications on structure functions in deep-inelastic scattering (DIS). Such imaging observables are expected to significantly enhance the impact of LHC data by providing more direct and effective constraints on designing the parametrization forms of flavor-separated nuclear modifications, which can facilitate the global nPDF fit. These intuitive images of $r^{\textrm{A}}_i(x,Q^2)$ can also help study the non-perturbative dynamics and binding mechanisms in nuclei~\cite{Alekhin:2022tip,Alekhin:2022uwc,Cloet:2012td,CLAS:2019vsb,Cocuzza:2021rfn,Dalal:2022zkg,nCTEQ:2023cpo}.

The feasibility of this method is validated through perturbative calculations at next-to-leading order using three parametrizations of nuclear PDFs. While the current precision of LHC measurements may not yet suffice to achieve a perfect image of $r^{\textrm{A}}_i(x,Q^2)$, the method proposed here opens new possibilities for optimizing future measurements. 

It should be emphasized that, both PDFs and their modification factors $r^{\textrm{A}}_i(x,Q^2)$ are not directly measurable, although this work demonstrates the ability of the imaging observables to reflect the underlying $r^{\textrm{A}}_i(x,Q^2)$ factors. In the context of global PDF extraction~\cite{Dulat:2015mca,Segarra:2020gtj,Hirai:2007sx,Eskola:2009uj,deFlorian:2011fp,Kovarik:2015cma,Wang:2016mzo,Walt:2019slu,Khanpour:2020zyu,Eskola:2021nhw,AbdulKhalek:2022fyi}, utilizing the data with better disentangled contributions from various values of $i$ and $x$ may help reduce the correlations and degeneracy of the fitting parameters and streamline uncertainty propagation analysis, resulting in enhanced computational efficiency and stability in the global extraction. Compared to traditional observables, the proposed imaging observables can provide somewhat preprocessed data with reduced entanglement of the contributions from PDF variables. Nevertheless, if such imaging observables share the common data source with the traditional measurements used in the analysis, we do not expect any statistically significant improvement in fitting the imaging observables relative to the analysis with usual measurements. 

It is also noteworthy that, this imaging technique can be generalized to study other processes at the LHC, such as the Drell-Yan process~\cite{Reimer:2007zza,CMS:2021ynu}, dijet production~\cite{CMS:2018jpl}, and $\gamma+$jet production~\cite{Neufeld:2010fj,Ma:2018tjv}. Its future applications in the EIC program, e.g., for dijet production~\cite{Guzey:2020zza,Zhang:2021tcc}, are also of significant interest. Importantly, this approach can help establish a framework for analysis and interpretation of the commonalities and distinctions in nuclear modification effects across diverse processes, spanning from $p$A to $e$A collisions. Furthermore, the study of nuclear effects at higher-twist, e.g. the parton multiple scattering~\cite{Kang:2013ufa,Ru:2019qvz,LHCb:2021vww}, may also benefit from such systematic comparisons, in which the leading-twist effects across various processes are effectively unified or mapped.

\begin{figure}[t]
\vspace{0.2cm}\includegraphics[width=2.8in]{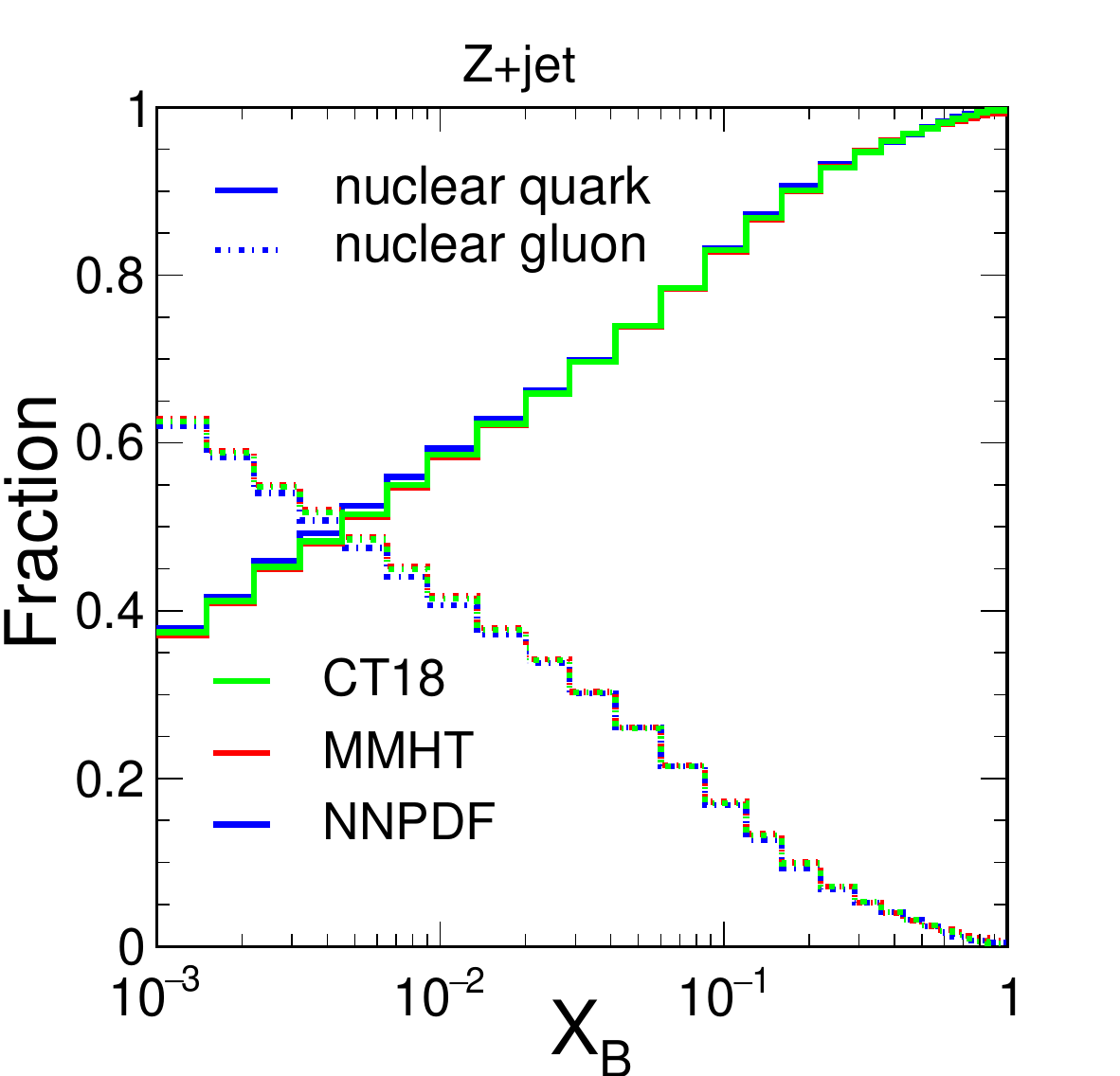}
\includegraphics[width=2.8in]{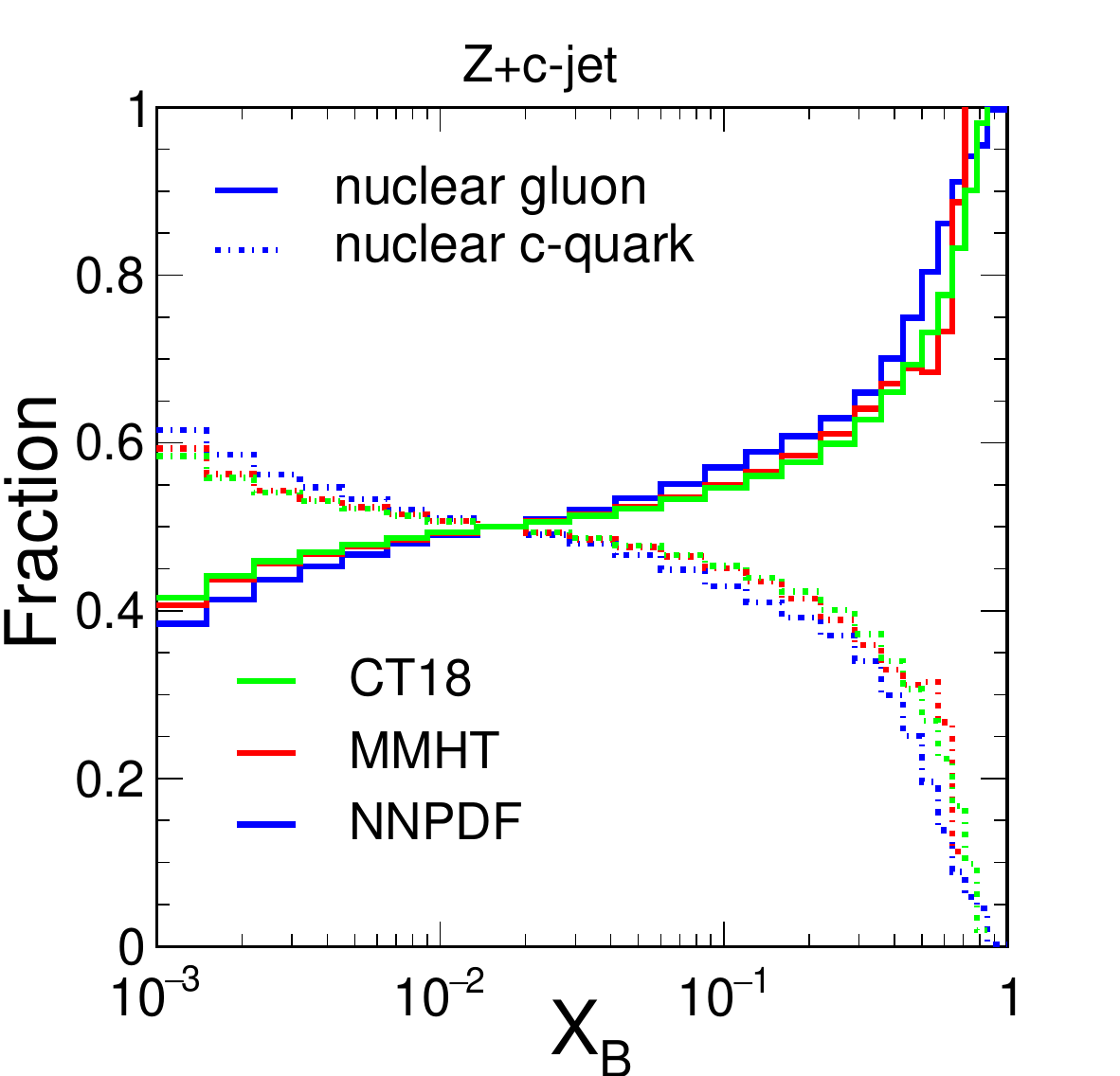}
\caption{Contribution fraction of certain nuclear parton flavors in $Z+$jet~(left) and $Z+c$-jet~(right) productions with three parametrizations of free-proton PDFs: CT18~\cite{Hou:2019efy}, MMHT~\cite{Harland-Lang:2019pla} and NNPDF4.0~\cite{Cruz-Martinez:2024cbz}.}
\label{fig:fraction}
\end{figure}

\begin{acknowledgements}
The authors would like to thank G.~Moore and X.-N.~Wang for helpful discussions. This research was supported in part by National Natural Science Foundation of China~(NSFC) under Project No. 12535010 and by Guangdong  Basic and Applied Basic Research Foundation under Project No.2022A1515110392.

\end{acknowledgements}

\begin{appendix}
\section{A test of impact from proton PDF parametrization}
\label{append:proton PDF}
To test the impact of the uncertainties in the relative compositions of various flavors in proton from PDF parametrization, we calculate the contribution fraction for certain nuclear parton flavors in $Z+$jet and $Z+c$-jet productions with three parametrizations of free-proton PDFs~(CT18~\cite{Hou:2019efy}, MMHT~\cite{Harland-Lang:2019pla} and NNPDF~\cite{Cruz-Martinez:2024cbz}). Results are plotted in Fig.\ref{fig:fraction}, and small distinctions among these results can be observed.

\end{appendix}


\end{document}